\documentclass[12pt,preprint]{aastex}

\shorttitle{VLBI astrometry of PSR J2222--0137}

\newcommand{\psrpi}{{PSR\large$\pi$}}
\newcommand{\Msun}{\ensuremath{M_{\odot}}}
\newcommand{\psr}{PSR J2222--0137}
\newcommand{\ibone}{J2222--0132}
\newcommand{\ibtwo}{J2221--0128}

\newcommand{\degrees}{\ensuremath{^\circ}}

\begin{document}

\title{VLBI astrometry of PSR J2222--0137: a pulsar distance measured to 0.4\% accuracy}

\author{
A.T. Deller\altaffilmark{1}, 
J. Boyles\altaffilmark{2}, 
D.R. Lorimer\altaffilmark{2}, 
V.M. Kaspi\altaffilmark{3},  
M.A. McLaughlin\altaffilmark{2}, 
S. Ransom\altaffilmark{4}, 
I.H. Stairs\altaffilmark{5}, 
K. Stovall\altaffilmark{6,7}}

\altaffiltext{1}{ASTRON, the Netherlands Institute for Radio Astronomy, Postbus 2, 7990 AA, Dwingeloo, The Netherlands}
\altaffiltext{2}{Department of Physics, West Virginia University, Morgantown, WV 26506, USA}
\altaffiltext{3}{Department of Physics, McGill University, 3600 University Street, Montreal, QC H3A 2T8, Canada}
\altaffiltext{4}{National Radio Astronomy Observatory, Charlottesville, VA 22903, USA}
\altaffiltext{5}{Department of Physics and Astronomy, University of British Columbia, 6224 Agricultural Road, Vancouver, BC V6T 1Z1, Canada}
\altaffiltext{6}{Center for Advanced Radio Astronomy and Department of Physics and Astronomy, University of Texas at Brownsville, Brownsville, Texas 78520}
\altaffiltext{7}{Department of Physics and Astronomy, University of Texas at San Antonio, San Antonio, Texas 78249}

\begin{abstract}
The binary pulsar J2222--0137 is an enigmatic system containing a partially recycled millisecond pulsar and a companion of unknown nature.  Whilst the low eccentricity of the system favors a white dwarf companion, an unusual double neutron star system is also a possibility, and optical observations will be able to distinguish between these possibilities. In order to allow the absolute luminosity (or upper limit) of the companion object to be properly calibrated, we undertook astrometric observations with the Very Long Baseline Array to constrain the system distance via a measurement of annual geometric parallax.  With these observations, we measure the parallax of the \psr\ system to be 3.742$^{+0.013}_{-0.016}$ milliarcseconds, yielding a distance of  267.3$^{+1.2}_{-0.9}$ pc, and measure the transverse velocity to be 57.1$^{+0.3}_{-0.2}$ km s$^{-1}$.  Fixing these parameters in the pulsar timing model made it possible to obtain a measurement of Shapiro delay and hence the system inclination, which shows that the system is nearly edge-on (sin $i = 0.9985 \pm 0.0005$).  Furthermore, we were able to detect the orbital motion of \psr\ in our VLBI observations and measure the longitude of ascending node $\Omega$.  The VLBI astrometry yields the most accurate distance obtained for a radio pulsar to date, and is furthermore the most accurate parallax for any radio source obtained at ``low" radio frequencies (below $\sim$5 GHz, where the ionosphere dominates the error budget).  Using the astrometric results, we show the companion to \psr\ will be easily detectable in deep optical observations if it is a white dwarf.   Finally, we discuss the implications of this measurement for future ultra--high--precision astrometry, in particular in support of pulsar timing arrays.
\end{abstract}


\section{Introduction}

PSR~J2222$-$0137 was discovered in the Green Bank Telescope 350-MHz drift-scan pulsar survey carried out in 2007 \citep{boyles13a,lynch13a}.  It has a observed spin period $P$ of 32.82~ms and spin period derivative $\dot{P}$ of $4.74\times10^{-20}$.  The dispersion measure is only 3.27~pc $\rm cm^{-3}$, which places the pulsar at a distance of roughly 300~pc assuming the NE2001 electron density model \citep{cordes02a}, although dispersion measure distances can exhibit large errors for individual objects \citep[e.g.][]{deller09b}.  PSR~J2222$-$0137 is in a low-eccentricity orbit ($e=0.00038$) with an orbital period of 2.4 days.  The spin period, low eccentricity, and small $\dot{P}$ indicate that PSR~J2222$-$0137 has been partially recycled.

Using the orbital parameters obtained from timing and assuming a pulsar mass of 1.35~$\Msun$ gives a minimum companion mass of 1.1~$\Msun$ \citep{boyles13a}.    Despite the relatively high minimum companion mass, the low orbital eccentricity argues against the likelihood that PSR~J2222$-$0137 is a member of a double neutron star (DNS) binary system.   For comparison, amongst known DNS systems the lowest measured eccentricity is around 0.09 \citep[for PSR J0737-3039;][]{lyne04a}, a factor of over 200 greater than \psr. The majority of DNS systems are expected to be born with a high eccentricity \citep{chaurasia05a}, so despite gravitational wave emission leading to circularization over time, such an extremely low eccentricity would be unexpected.  A relatively heavy CO white dwarf companion is the alternative explanation, which would make PSR~J2222$-$0137 an ``intermediate-mass binary pulsar" \citep[e.g.,][]{camilo01a}.

Characterizing the \psr\ system and distinguishing between the possible evolutionary pathways will require multiwavelength data which can be reliably interpreted.  This demands an accurate distance to the system, in order to convert observed flux densities to absolute luminosities.  Very Long Baseline Interferometry (VLBI) can provide astrometric accuracies on the order of tens of microarcseconds, sufficient to measure distances accurately out to a range of $\sim$10 kpc through the measurement of annual geometric parallax.  The Very Long Baseline Array (VLBA) has demonstrated an outstanding capability for precision astrometry, having been used to map a variety of Galactic objects such as pulsars, masers and low--mass protostars with exquisite precision \citep[e.g.][]{chatterjee09a,reid09a,loinard07a}.  At the relatively low radio frequencies usually required for pulsar observations ($\lesssim$5 GHz, where the ionosphere dominates error budgets) the ability of the VLBA to make use of ``in--beam" calibrators for the majority of targets gives it a particular advantage \citep{chatterjee09a}.  Accordingly, we undertook an astrometric campaign on \psr\ using the VLBA to determine its distance.

\section{Observations and data reduction}

We observed \psr\ a total of 8 times with the VLBA between July 2010 and June 2012.  Each observation had a duration of 2 hours, and used the source J2218--0335 as the primary calibrator, which is separated from \psr\ by 2.1\degrees.  In order to maximize the astrometric accuracy, our first observation focused on the identification of a suitable in--beam calibrator, which can be observed contemporaneously with the target and reduces the spatial and temporal interpolation of calibration solutions.  The use of an in--beam calibrator is particularly important at the low frequencies typical for pulsar astrometry, since astrometric precision is then dominated by fluctuations in the ionosphere which are difficult to model and remove \citep[e.g.,][]{deller12b,chatterjee09a}.

\begin{figure}
\begin{center}
\includegraphics*[height=0.45\textwidth]{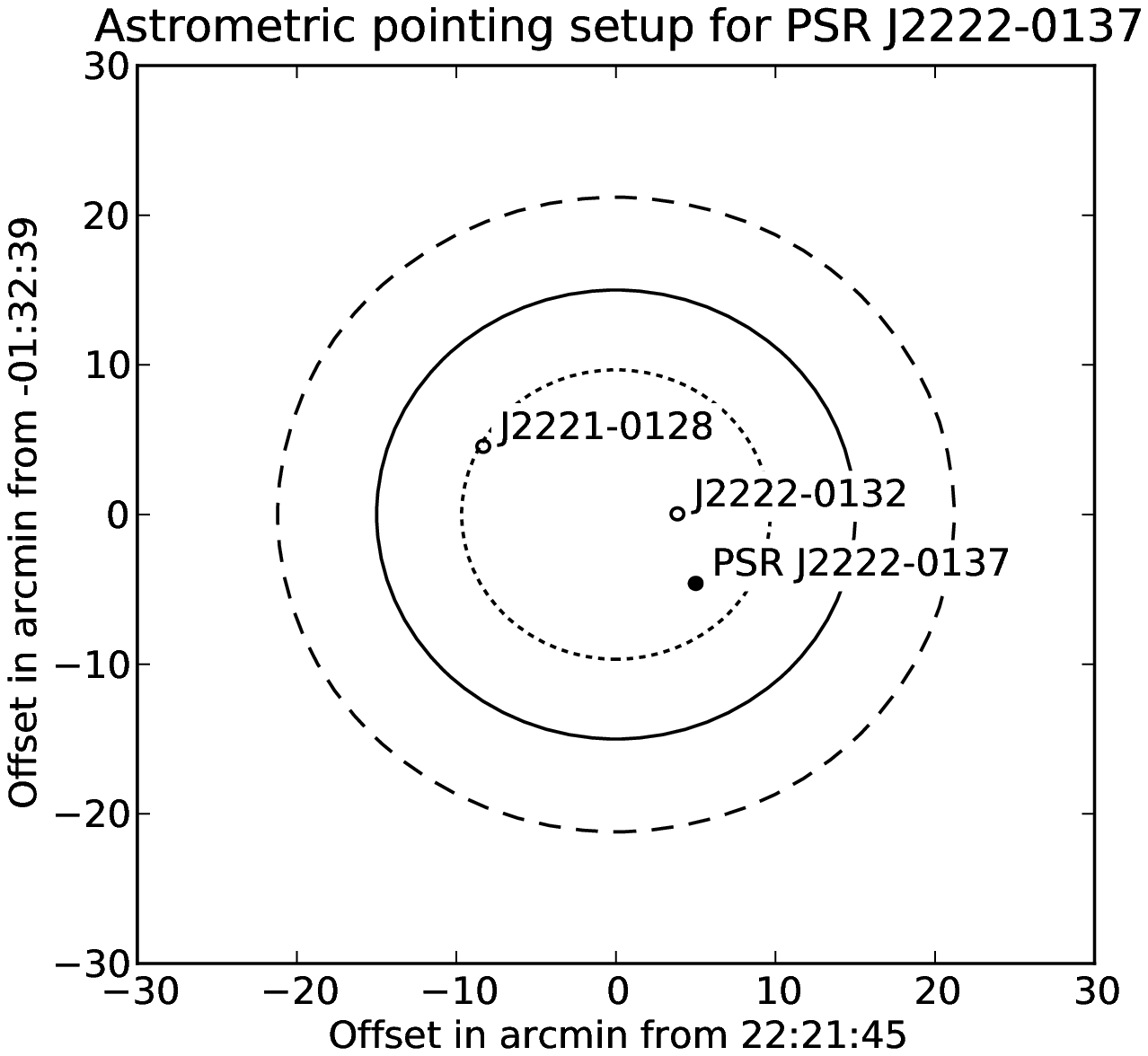}
\caption{The pointing layout for astrometric observations.  All of the sources lie within the inner dotted line which shows the 75\% response point of the primary beam.  The 50\% response point and 25\% response point of the beam are shown as a solid and dashed line, respectively, for scale.}
\label{fig:pointing}
\end{center}
\end{figure}

\begin{figure}
\begin{center}
\includegraphics*[height=0.45\textwidth, angle=270,clip=]{f2a.ps}
\includegraphics*[height=0.45\textwidth, angle=270,clip=]{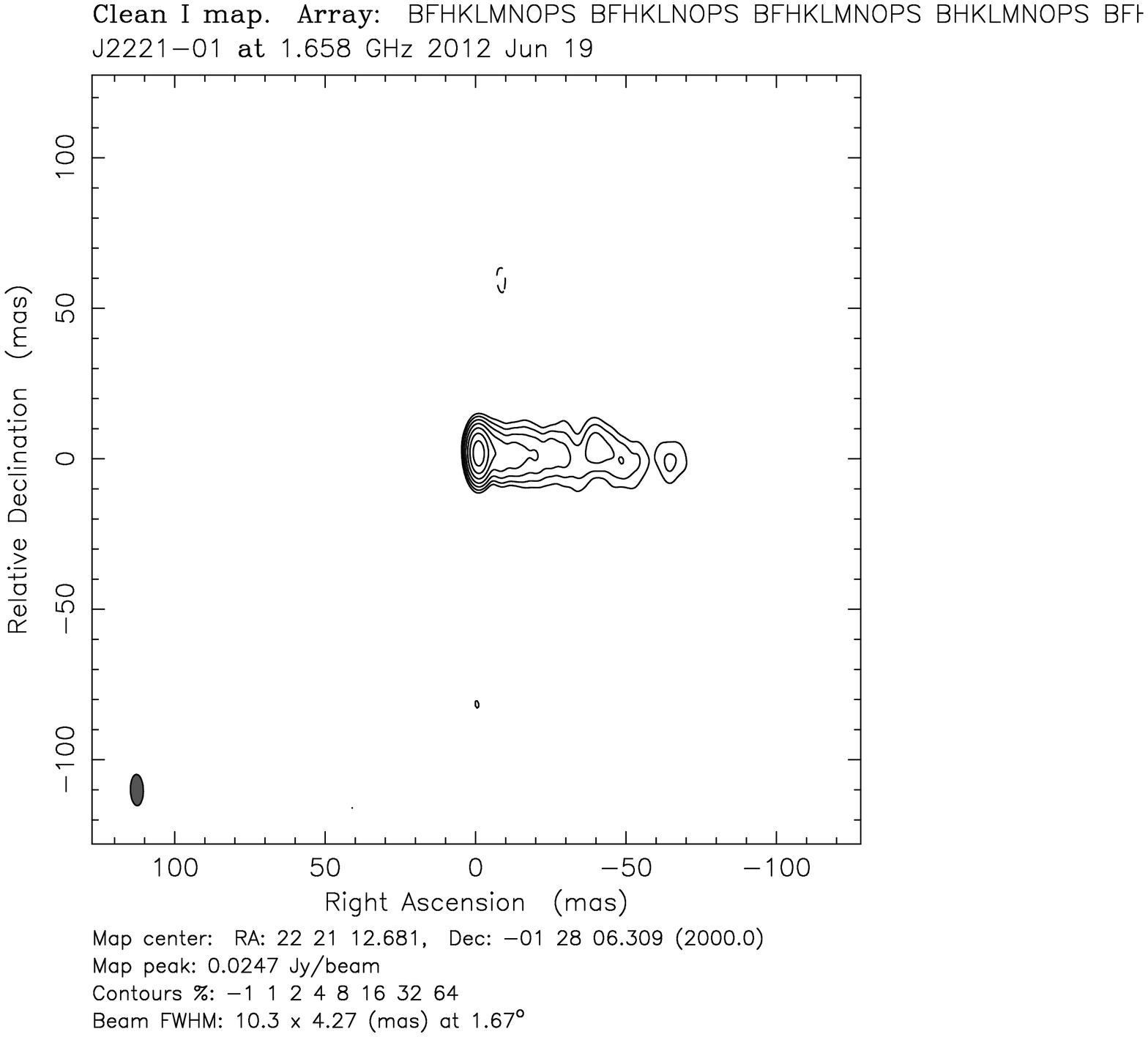}
\caption{The two in--beam calibrator sources, imaged using all astrometric epochs (center frequency 1650 MHz) combined.  Contours increase by factors of two.  {\em (Left)} J2222--0132; peak flux 16 mJy/beam, lowest contour beginning at 0.5\% of the peak. The faint extended structure to the south--east is real and included in the model.  {\em (Right)} J2221--0128; peak flux 25 mJy/beam, lowest contour 1\% of the peak.}
\label{fig:inbeams}
\end{center}
\end{figure}

This initial search observation was conducted at 1.4 GHz and targeted all sources from the FIRST survey \citep{becker95a} which fell within the primary beam of the VLBA, using the multifield correlation mode of the DiFX software correlator \citep{deller11a} and the observation setup described in \citet{deller11b}.  Of the 30 sources targeted, 11 were detected with peak flux densities ranging from 0.3 to 13 mJy/beam.  Although \psr\ was detected in this first epoch, the position obtained was not used in the subsequent astrometric analysis described in Section~\ref{sec:fits}, because the pointing center and (most importantly) frequency setup differed substantially from the later epochs.  Based on proximity, compactness and brightness, FIRST J222201--013236 (hereafter J2222--0132) was chosen to be the primary in--beam calibrator.  The pointing center for scans on the target was placed at right ascension 22:21:45.95, declination --01:32:39.67, which placed \psr\ and J2222--0132 near the pointing center but also allowed FIRST J222112--012806 (hereafter J2221--0128) to fall within the VLBA primary beam.  \ibtwo\ is also bright, but less compact and further from \psr.  Table~\ref{tab:calibrators} summarizes the calibrator positions, and Figure~\ref{fig:pointing} shows the layout on the sky.  Images of the two in--beam calibrators are shown in Figure~\ref{fig:inbeams}.

\begin{deluxetable}{lllc}
\tabletypesize{\small}
\tablecaption{Calibrator sources}
\tablewidth{0pt}
\tablehead{
\colhead{Source name} & \colhead{Right ascension} & \colhead{Declination} & \colhead{Peak flux density (mJy/beam)}
}
\startdata
J2218--0335\tablenotemark{A}	&  22:18:52.037725	&	$-$03:35:36.87963 		& 1480  \\
J2222--0132	&  22:22:01.373502	&	$-$01:32:36.98196		& 15 \\
J2221--0128	&  22:21:12.681147	&	$-$01:28:06.31288		& 21 
\enddata
\tablenotetext{A}{The absolute position error of J2218--0335 is 0.1 mas in each coordinate -- this error also dominates the absolute position error of J2222--0132 and J2221--0128. The position of J2218--0335 was taken from the rfc2011d catalog (\texttt{http://astrogeo.org/rfc/}).}
\label{tab:calibrators}
\end{deluxetable}

The observing setup for the remaining 7 astrometric epochs was as follows.  Left and right polarizations were sampled in 4 subbands, each of width 16 MHz, with a total data rate of 512 Mbps/antenna.  The bands were placed adjacent to one another and spanned the frequency range  1626.49 -- 1690.49 MHz.  The final six observations were clustered in pairs, with each pair sampling close to the time of parallax extrema.  A phase reference cycle time of 7 minutes was used, with a 1 minute scan on the external phase reference source J2218--0335 followed by 6 minutes on the target pointing.  In total, 90 minutes of time was obtained on--source for the target per observation, with a typical 1$\sigma$ image rms of 65 $\mu$Jy/beam.  For each epoch, 3 correlation passes were made using the DiFX software correlator \citep{deller07a}.  All correlator passes used an averaging time of 2 seconds and a frequency resolution of 0.5 MHz.  The first two correlator passes did not use any pulsar gating and used the positions of J2222--0132 and J2221--0128 for the target pointing.  The third correlator pass used a simple pulsar gate with width of 4\% of the pulsar period (which encompassed the pulse down to the 10\% level -- the pulse profile can be seen in \citealt{boyles13a}), providing a factor of 5 gain in sensitivity.  The pulsar ephemeris was updated during the course of the observations, using the timing observations presented in \citet{boyles13a}.

The visibility data produced by the correlator were reduced using AIPS\footnote{http://www.aips.nrao.edu/}, utilizing standard scripts based on the ParselTongue package \citep{kettenis06a}. After loading the data and flagging known bad data, the logged system temperature data was used to calibrate visibility amplitudes.  Significant radio--frequency interference (RFI) was seen in the highest frequency subband (1674.49 -- 1690.49 MHz), which led to unreliable system temperature information and calibration solutions for many stations.  Additionally, the ``Mark5A" recording system (which will soon be retired as part of the VLBA sensitivity upgrade) at some VLBA stations exhibits delay jumps at unpredictable intervals (with a typical timescale of tens of minutes) in its 7th recording channel when recording at 512 Mbps.  In our observing setup, the 7th recording channel is the R polarization of the highest frequency subband.  The combination of RFI and delay jumps rendered this subband unsuitable for precise astrometry, and so we flagged and discarded this subband in all epochs, reducing our effective bandwidth to 48 MHz.

Calibration based on global ionospheric models was applied using the AIPS task TECOR.  Subsequently, the delay and bandpass were calibrated for each subband independently using J2218--0335, and the amplitude calibration was refined with one round of self--calibration on the same source.  At this time, the data were split and averaged in frequency to a single point per subband, and all future calibration was performed using this averaged data in Stokes I.   Phase--only corrections were generated from the primary in--beam J2222--0132 with a solution interval of 1 minute and applied to the other sources in the target field (\psr\ and J2221--0128).  For each calibrator source (J2222--0335, J2222--0132, and J2221--0128), a combined model was formed based on the data from all epochs, and all calibration steps made use of the appropriate model.  Despite the relatively narrow fractional bandwidth, the effect of different spectral indices in the two spatial components of J2222--0132 could clearly be seen, and so for this source a model which included components with a spectral slope was generated (the image in Figure~\ref{fig:inbeams} shows the central frequency).  For all sources, the models were not permitted to vary between epochs.  No correction was made for the motion of \psr\ during the observation (over the two hours, the source moves by $\sim$15 $\mu$as, insignificant compared to the measurement errors).

Once all calibration was applied, the visibility data for each source from the target field was written to disk and and imaged using difmap \citep{shepherd97a} with natural weighting.  A  ``combined" Stokes I image was formed utilizing all subbands; each 16--MHz subband was also imaged in Stokes I separately.  In each image, a single gaussian component was fitted using the AIPS task JMFIT, and the position and errors were used in the following astrometric analysis.  Since J2221--0128 has complicated resolved structure, a gaussian fit in the image plane could be affected by beam--shape effects in different epochs (when equipment failure causing the absence of different antennas changes the $uv$ coverage).  Accordingly, for J2221--0128, we divided the $uv$ data by our average model, a procedure which will (given a perfect model) transform the image into a point source at the phase center, and avoid the problem of beam--shapes.  Since \psr\ is already point--like, no such step was required for this target.

\section{Astrometric fits and results}
\label{sec:fits}

\psr\ is a member of a select group of binary pulsars which are close enough to the Earth and have sufficiently long orbital periods that orbital motion of the pulsar is detectable.  This affords the rare opportunity to measure the longitude of ascending node $\Omega$, which has only been achieved via pulsar timing for a couple of millisecond pulsars \citep{splaver05a,verbiest08a}.  Such a measurement has not been made before using VLBI, although the currently--underway \psrpi\ program \citep{deller11b} will likely make similar measurements for PSR~J0823+0159, PSR~J1022+1001 and PSR~J2145--0750.  From pulsar timing, the orbital period $P_b$, eccentricity $e$, projected semi--major axis $a$ sin $i$ and argument of periastron $\omega$ are known \citep{boyles13a}.  Accordingly, both the inclination $i$ and longitude of ascending node $\Omega$ remain to be determined.  We note that in \citet{boyles13a} and in our results below, the definition of $\omega$ follows standard pulsar timing practice and is measured from the longitude of descending node, rather than the longitude of ascending node as is customary in other areas of astronomy.  $\Omega$ follows standard practice and is measured from north towards east.

Initially, we fitted the VLBI positions to only the traditional 5 astrometric parameters (reference right ascension $\alpha_0$, reference declination $\delta_0$, proper motions $\mu_\alpha$ and $\mu_\delta$ and parallax $\pi$), ignoring the effect of orbital motion.  Fitting to the 7 positions obtained from the combined image at each epoch, we obtain the values shown in the left column of Table~\ref{tab:fits}.  These values were then fixed in the pulsar timing model for \psr\ and the pulsar timing dataset was refitted.  Previously, covariances with parameters such as proper motion and position had prevented a measurement of the Shapiro delay for \psr.  With the astrometric parameters fixed, a significant measurement of the Shapiro delay for \psr\ was obtained, which in turn yields the inclination function sin $i$ = 0.9985 $\pm$ 0.0005.  Following standard pulsar timing practice, the error reported here is twice the formal timing error reported by \verb+tempo+.  This gives an inclination $i$ of 86.9\degr $\pm$ 0.5\degr\ or 93.1 $\pm$ 0.5\degr.  A full analysis of the improved timing model for \psr\ will be presented in a forthcoming paper (Boyles et al., 2013, in prep.)

\begin{figure}
\begin{center}
\includegraphics[width=0.85\textwidth]{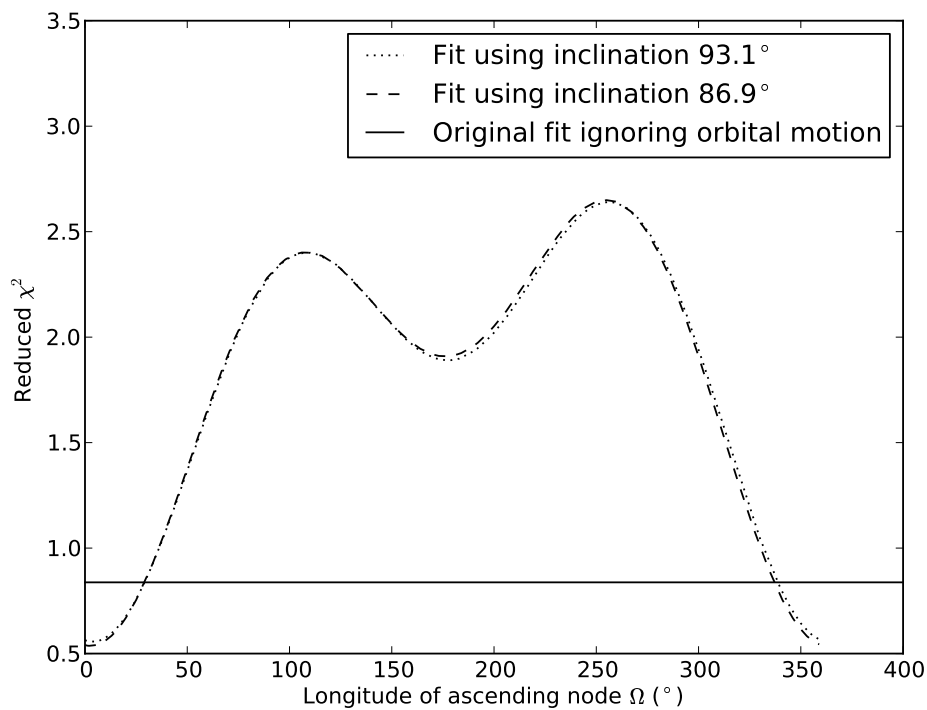}
\end{center}
\caption{The fit to longitude of ascending node $\Omega$ for \psr.  Very little difference is seen between the two possible inclination values, since the inclination is so close to 90\degr, so the VLBI observations are unable to distinguish between these two possibilities.  The best--fit value for $\Omega$ is 2\degr; this gives a considerably better fit than when the orbital motion is neglected entirely.}
\label{fig:orbfit}
\end{figure}

Subsequently, we performed a grid search for $\Omega$ between 0 and 360\degr\ with an interval of 1 degree, allowing $i$ to take the values 86.9\degr\ or 93.1\degr.  For each trial, we calculated the positional offset due to orbital motion at each astrometric epoch for the given value of $\Omega$ and $i$, and subtracted this offset, yielding a pulsar position corrected to the orbit center.  These corrected positions were then fitted for $\alpha_0$, $\delta_0$, $\mu_\alpha$, $\mu_\delta$, and $\pi$, and the resultant reduced $\chi^2$ was noted (where the reduced $\chi^2$ was calculated accounting for the changed number of degrees of freedom).  This yields the curves shown in Figure~\ref{fig:orbfit}, which shows that the astrometric results are unable to significantly distinguish between the two possible values of $i$, but that $\Omega$ can be clearly determined, with a best--fit value of 2\degr.  The best fit values for $\alpha_0$, $\delta_0$, $\mu_\alpha$, $\mu_\delta$, and $\pi$ when accounting for orbital motion are shown in the center column of Table~\ref{tab:fits}.  

\begin{figure}
\begin{center}
\begin{tabular}{cc}
\includegraphics[width=0.45\textwidth]{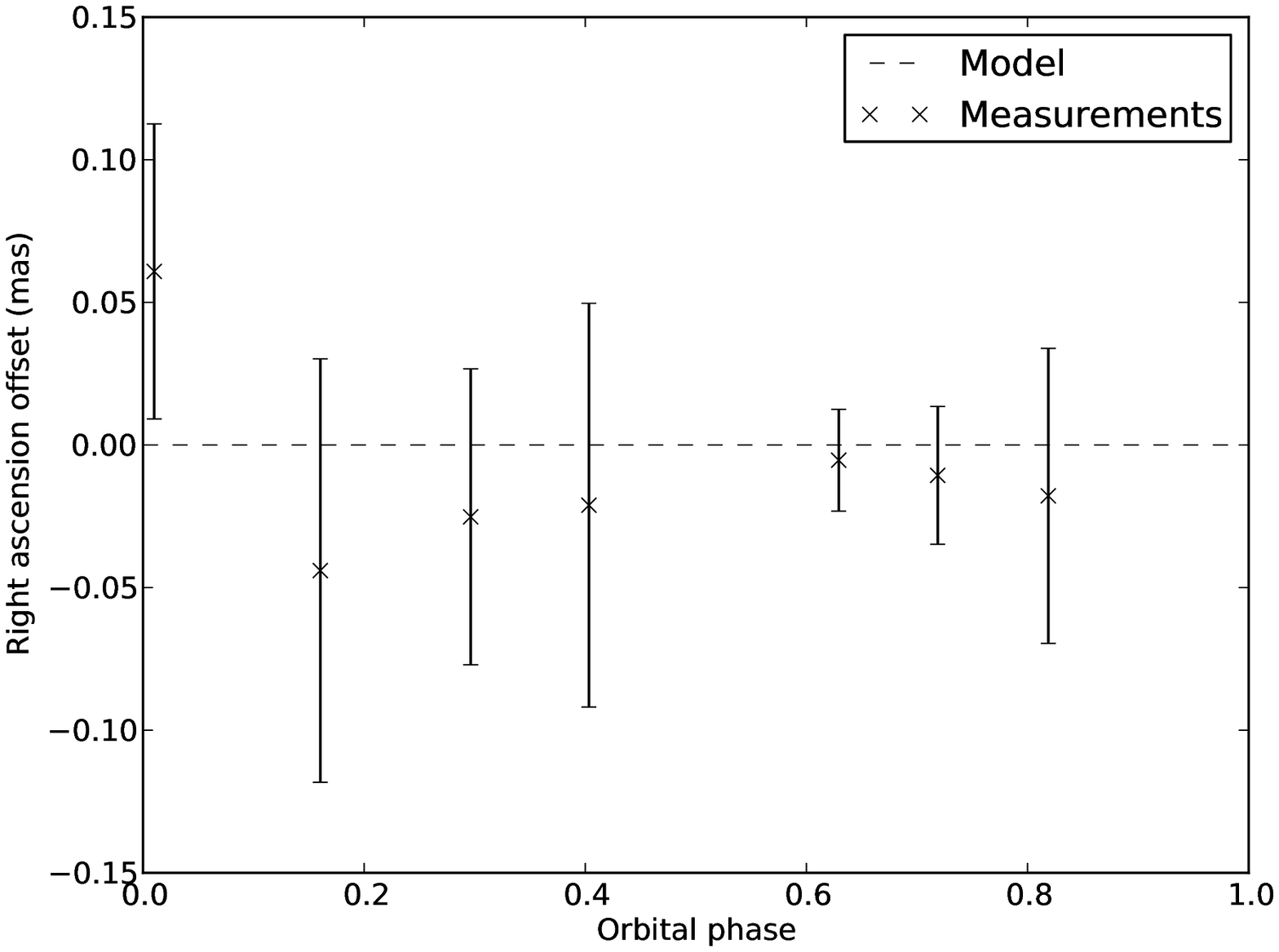} &
\includegraphics[width=0.45\textwidth]{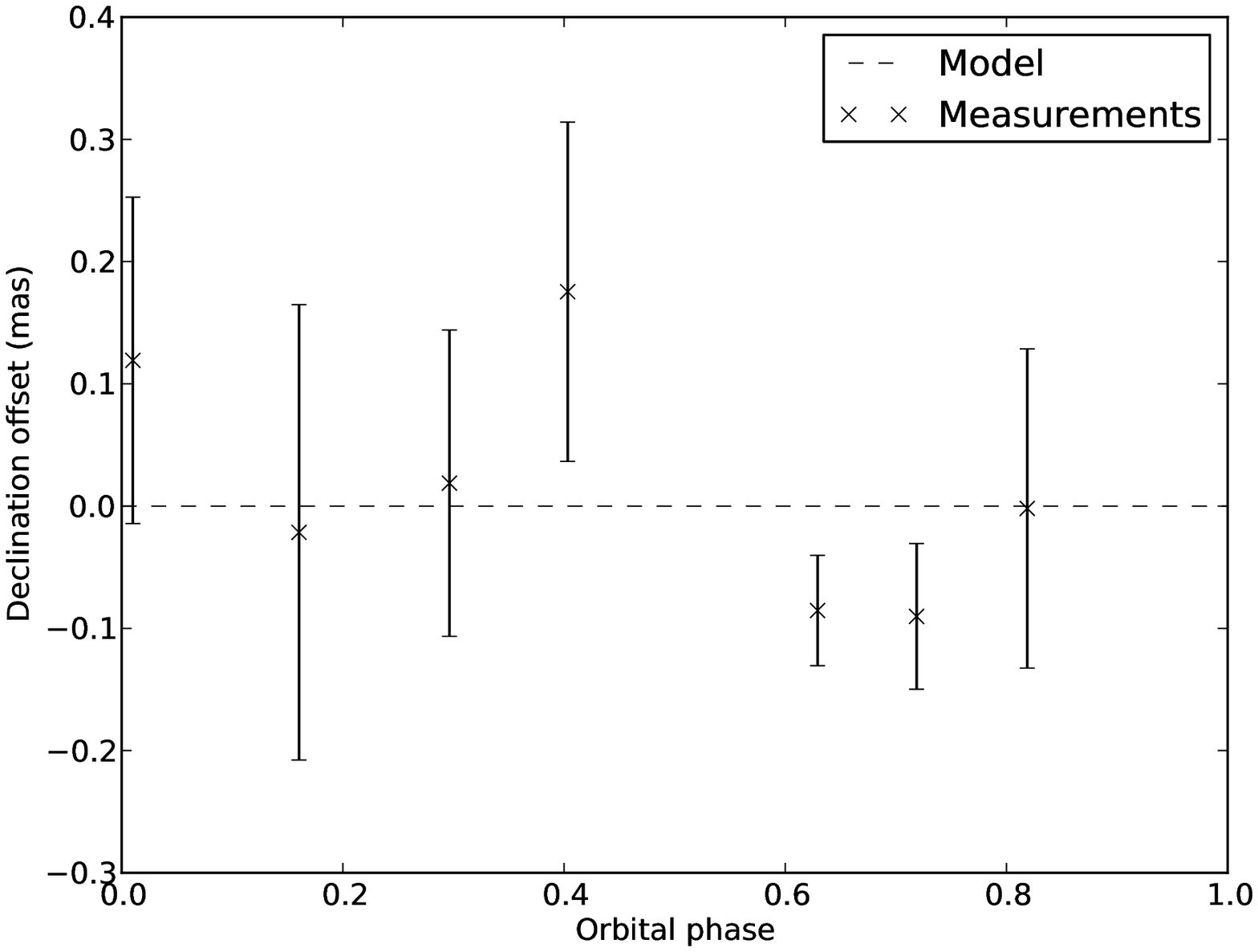} \\
\includegraphics[width=0.45\textwidth]{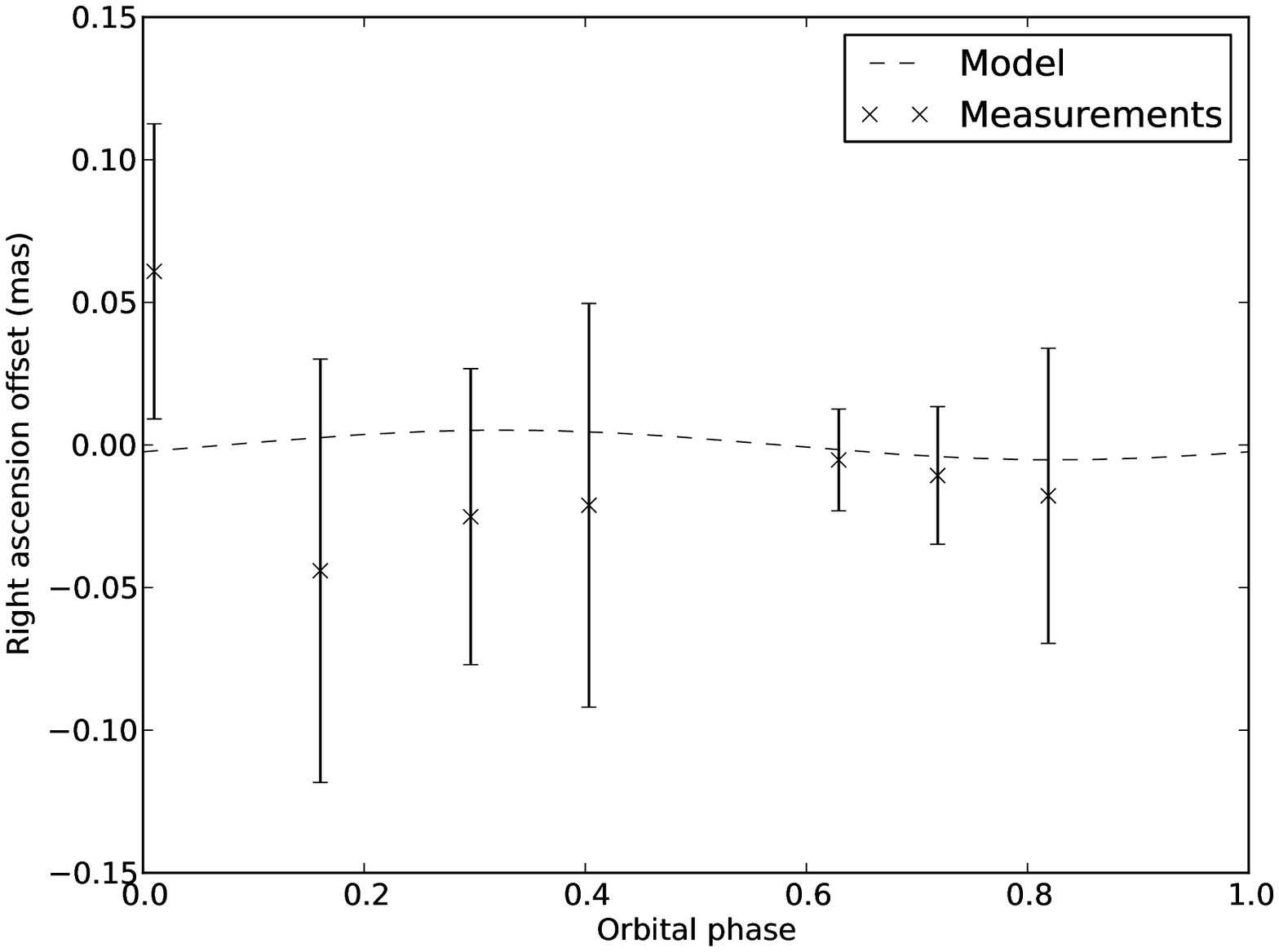} &
\includegraphics[width=0.45\textwidth]{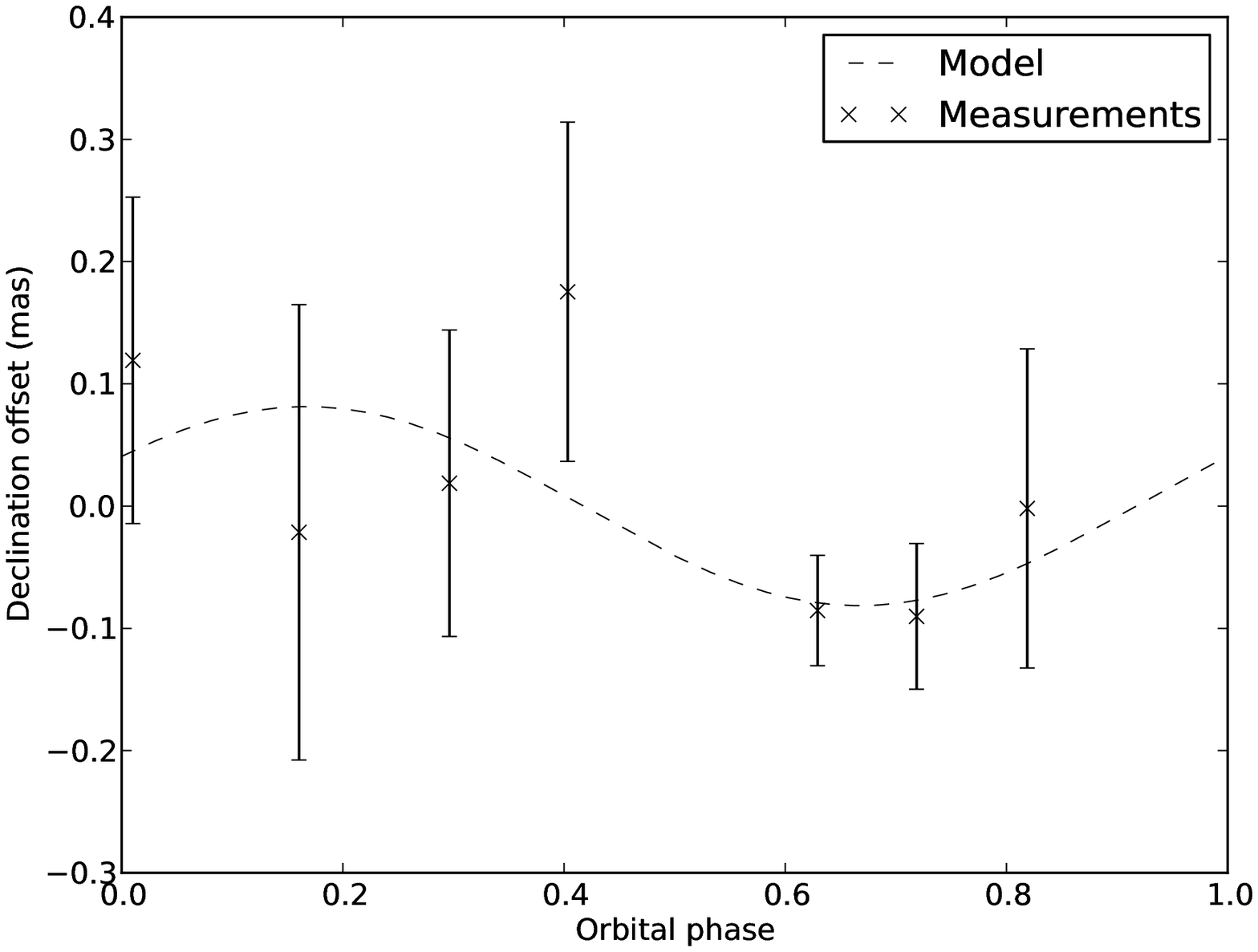}
\end{tabular}
\end{center}
\caption{Measured pulsar offset relative to the center of mass (left panels showing right ascension, right panels showing declination), plotted against orbital phase. The dashed line shows the fitted motion of the pulsar relative to the center of mass in the assumed model.  {\bf Top:} Results from the original astrometric fit, which assumed no orbital motion and therefore a pulsar position coincident with the center of mass (hence the dashed line is constant at 0).  {\bf Bottom:} Results obtained when the positions at each epoch are corrected for the best--fit orbital motion ($\Omega=5$\degr) before fitting the remaining 5 astrometric parameters.  The improvement (particularly in declination, where the effect of the orbital motion is concentrated) is noticeable.}
\label{fig:orbitoffsets}
\end{figure}

From Table~\ref{tab:fits}, it is immediately apparent that including or neglecting the orbital motion does not make a substantial impact on the other astrometric parameters.  This is largely due to the fact that the transverse orbital motion is largely confined to the declination axis, where the precision of the VLBI measurements are lower due to the VLBA beam shape.  This is also shown more clearly in Figure~\ref{fig:orbitoffsets}, which plots the residual offsets in right ascension and declination after subtracting the best fit values for $\alpha_0$, $\delta_0$, $\mu_\alpha$, $\mu_\delta$, and $\pi$.  The top panels show the results obtained when there is assumed to be no orbital motion, while the bottom panels show the results obtained when accounting for the orbital motion using the best--fit value for $\Omega$.

Typically, the approach taken above (of fitting the 5 astrometric parameters to a single position measurement for each epoch) will underestimate the error on each epoch, because it fails to account for systematic errors due to the ionosphere.  Therefore, such ``raw" pulsar astrometric fits typically have a reduced $\chi^2$ exceeding 1.0 \citep[see e.g.,][]{deller12b,deller09b}, but in this case the reduced $\chi^2$ of the fit is less than 1.0.  The implied negligible contribution of systematics in this case can be attributed to the small angular separation between \psr\ and J2222--0132 and the relative brightness of J2222--0132, which allows solutions on short timescales.  It also implies that the core position of the calibrator source J2222--0132 is stable at the level of tens of microarcseconds over a period of two years.

However, whilst the expectation value for the reduced $\chi^2$ of an astrometric fit is 1.0 if the measurement errors are accurate, the presence of measurement noise means that for any given sample of measurements -- even if the measurement errors are known perfectly -- a valid fit might obtain a reduced $\chi^2$ slightly less than or slightly greater than 1.0.   This effect is particularly severe when the number of degrees of freedom is relatively small, as is typically the case for astrometric observations.  Accordingly, a useful cross--check is a bootstrap test, which has been widely used in past pulsar astrometry projects \citep{chatterjee09a,deller12b,moldon12a}, since it can be used to estimate errors on fitted parameters when the underlying measurement errors are poorly known \citep{efron91a}.  Bootstrapping involves creating a large number of test datasets, where each dataset is constructed by sampling with replacement from the pool of measured astrometric positions.  The astrometric observables are fitted once from each test dataset and the large sample of tests is used to build a histogram of the fitted values for each observable.  In addition to cross-checking the errors on the 5 regular astrometric observables, this bootstrap fit allows a useful estimate of the error on $\Omega$, which would otherwise be difficult to obtain.

For the bootstrap test, we used the positions obtained from the images of single subbands ($7\times3 = 21$ measurements in total). Using the combined measurements from each epoch yields a sample of just 7 measurements, which is too small to make effective use of the bootstrap technique.  We made 10,000 trials, where in each trial we again performed a grid search for $\Omega$ between 0 and 360\degr\ with an interval of 1 degree, for a total of 3.6 million fits.  From each of the 10,000 trials, we recorded the best--fit $\alpha_0$, $\delta_0$, $\mu_\alpha$, $\mu_\delta$, $\pi$, and $\Omega$, and then constructed a cumulative probability histogram for each parameter from which we obtained the most compact 67\% probability interval.  The results are shown in the rightmost column of Table~\ref{tab:fits}.  The bootstrap test shows that we are able to measure the value of $\Omega$ with a 1$\sigma$ accuracy of $\sim$20\degr.
The agreement in the other 5 fitted values when bootstrapping compared to the simple fit is extremely good, with all values overlapping to well within 1$\sigma$.  


\begin{deluxetable}{lrrr}
\tabletypesize{\scriptsize}
\tablecaption{Fitted and derived astrometric parameters for \psr.}
\tablewidth{0pt}
\tablehead{
\colhead{Parameter} & \colhead{Standard fit} & \colhead{Standard fit} & \colhead{Bootstrap fit\tablenotemark{a}} \\
 & \colhead{(Orbital motion ignored)} & \colhead{(Orbital motion corrected)} & \colhead{(Orbital motion corrected)} 
}
\startdata
$\alpha_0$ (J2000)\tablenotemark{b}
							& 22:22:05.969101(1)
							& 22:22:05.969101(1)
							& 22:22:05.969101(1) \\
$\delta_0$ (J2000)\tablenotemark{b}
							& $-$01:37:15.72447(3)
							& $-$01:37:15.72444(3)
							& $-$01:37:15.72441(4)  \\
Position epoch (MJD)			& 55743 & 55743 & 55743 \\
$\mu_{\alpha}$	(mas yr$^{-1}$)		& 44.72 $\pm$ 0.02
							& 44.73 $\pm$ 0.02
							& 44.73 $\pm$ 0.02   \\
$\mu_{\delta}$	(mas yr$^{-1}$)		& $-$5.64 $\pm$ 0.06	
							& $-$5.68 $\pm$ 0.05
							& $-$5.68 $\pm$ 0.06   \\
Parallax (mas)	 				& 3.743 $\pm$ 0.010	
							& 3.742 $\pm$ 0.010
							& 3.742$^{+0.013}_{-0.016}$  \\
Distance (pc)					& 267.2 $\pm$ 0.7 
							& 267.3 $\pm$ 0.7
							& 267.3$^{+1.2}_{-0.9}$ \\
$v_{\mathrm T}$ (km s$^{-1}$)		& 57.1 $\pm$ 0.2
							& 57.1 $\pm$ 0.2
							& 57.1$^{+0.3}_{-0.2}$ \\
$\Omega$ (\degr)				& -- 
							& 2\tablenotemark{c}
							& 5$^{+15}_{-20}$ \\
Reduced $\chi^2$				& 0.84 & 0.53 & n/a
\enddata
\tablenotetext{a}{Values from the combined bootstrap fit (including the solution for $\Omega$) are used in the analysis.}
\tablenotetext{b}{The errors quoted here are from the astrometric fit only and do not include the $\sim$0.1 mas position uncertainty transferred from the in--beam calibrator's absolute position.}
\tablenotetext{c}{No attempt was made to estimate an error for $\Omega$ based on the standard astrometric fit.}
\label{tab:fits}
\end{deluxetable}

 \begin{figure}
\begin{center}
\begin{tabular}{c}
\includegraphics[width=0.45\textwidth]{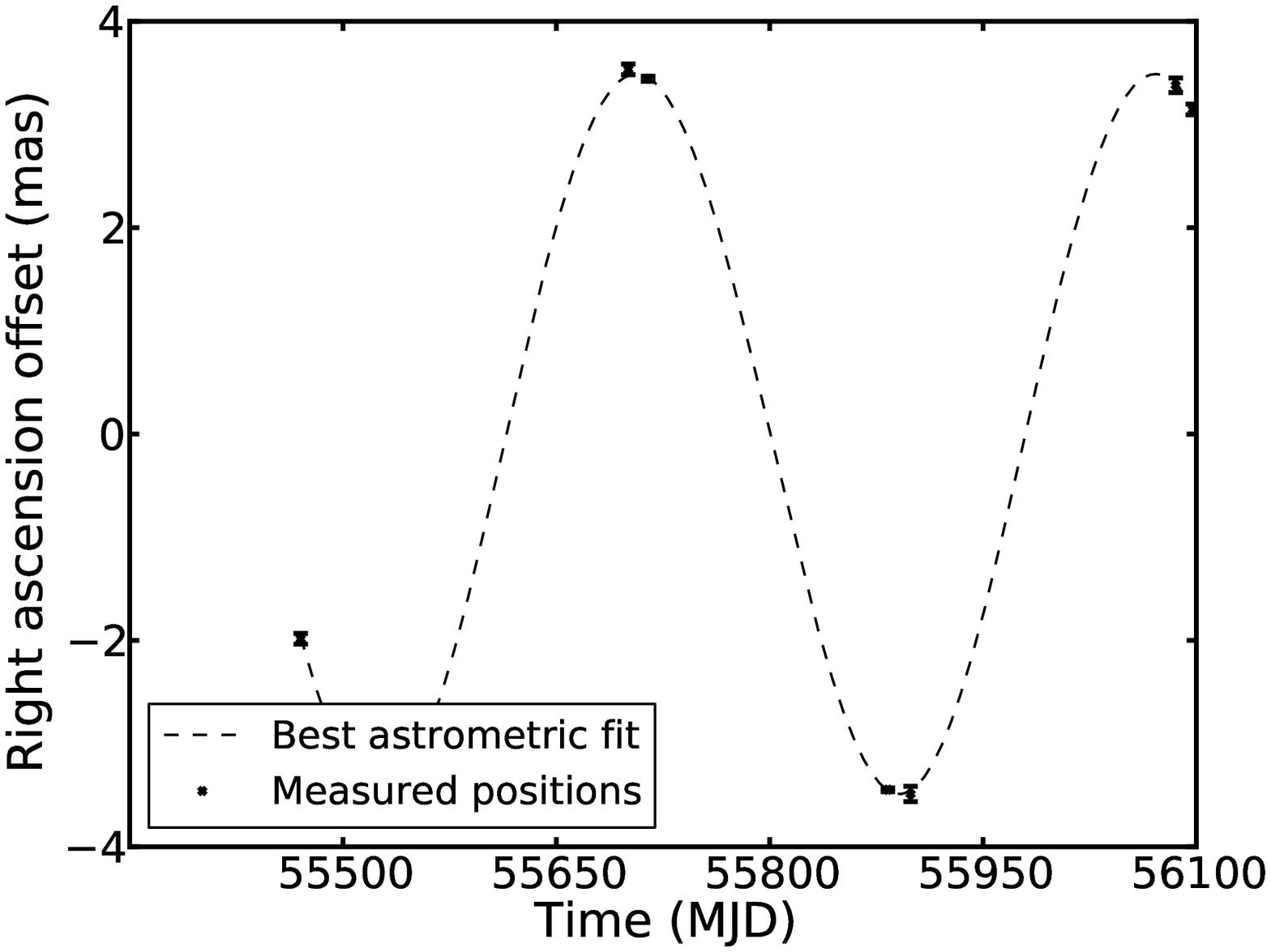} \\
\includegraphics[width=0.45\textwidth]{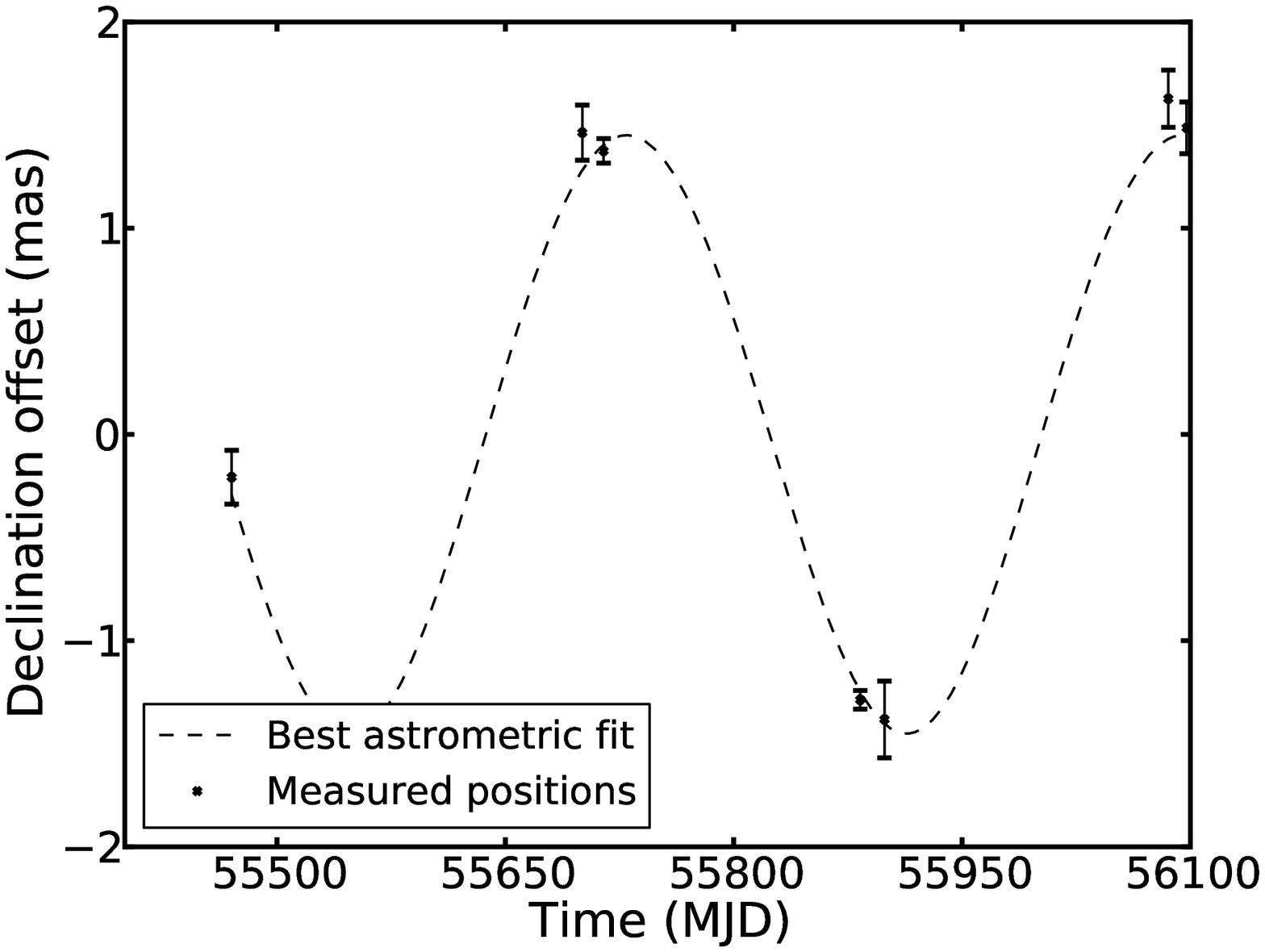} 
\end{tabular}
\end{center}
\caption{The astrometric fit to the positions obtained for \psr, highlighting the parallax signature.  The top panel shows offset from the reference position (at MJD 55743) in right ascension after the subtraction of the best--fit proper motion; the bottom panel shows the same for declination.  Both the amplitude of the parallax signature and the precision of the VLBA measurement are greater in the right ascension coordinate, which is the reason why the epochs are grouped near the parallax extrema in right ascension.}
\label{fig:psrfit}
\end{figure}

The error intervals themselves are also comparable, although the bootstrap errors are generally slightly more conservative.   Partly, this is because of the covariance between $\Omega$\ and parallax, which is not accounted for in the simple fit.  However, an additional concern noted by \citet{deller12b} is that bootstrapping in pulsar astrometry suffers from the drawback that constructing a sufficiently large sample size requires the use of positions obtained from images of single subbands, where the position errors can become non--linear at low S/N.  For our observations, the  epochs where the pulsar was faintest had a S/N of $\sim$15 in the single--band images, low enough that this effect may be present, which would lead to the bootstrap method overestimating the errors slightly.  However, since this method is the only way to obtain a useful estimate of the error of $\Omega$, we use the values and errors from the bootstrap fit in the analysis below. 
To highlight the parallax measurement, the combined image position measurements and the astrometric fit (after the subtraction of proper motion and orbital motion) are shown in Figure~\ref{fig:psrfit}.

This astrometric measurement is groundbreaking in several respects.  It is the most accurate pulsar parallax obtained to date, with an error $\sim$30\% lower than PSR J1543--0929 \citep{chatterjee09a}.  It is also the most accurate pulsar distance, with an error $\sim$30\% lower than PSR J0437--4715 \citep{deller08b}.  
Finally, it is the first pulsar for which $\Omega$ has been directly measured using VLBI (for PSR B1259--63, \citealt{moldon11a} inferred a value for $\Omega$ based on morphological measurements, but did not make a direct measurement).
Despite the challenges of astrometry at 1.6 GHz compared to higher frequency observations (lower resolution, greatly increased ionospheric effects), the accuracy approaches those obtained with maser measurements at 22 GHz \citep[7 $\mu$as;][]{nagayama11a}.  It suggests that extremely high precision astrometry should be possible even at low frequency with the continued evolution of VLBI instrumentation.  The implications for future astrometric studies and some possible applications of extremely high precision pulsar distance measurements are discussed in Section~\ref{sec:general}.
 
 \section{Implications for \psr}
 
The distance of 267 pc places the pulsar around 15\% closer than estimates based on its dispersion measure \citep[312 pc using the NE2001 electron density distribution model;][]{cordes02a}.  A discrepancy at this level is consistent with the predictive power of these models.  The transverse velocity of \psr\ is $57.1\pm0.2$ km s$^{-1}$, typical of a recycled pulsar in a binary system.  Correction for peculiar solar motion and Galactic rotation using a flat rotation curve and the current IAU recommended rotation constants (R$_{0}$ = 8.5 kpc, $\Theta_{0}$ = 220\,km\,s$^{-1}$) alters this value slightly to $46.6 \pm 0.2$ km s$^{-1}$.  The nearer--than--expected distance coupled with the lack of an identified optical companion means that the optical emission from the companion to \psr\ must be very faint \citep{boyles11a}.  At a distance of 267 pc, the presence of even an extremely old and cold massive white dwarf will be easily detectable with a large ground based telescope.  For example, at a temperature of 5000 K (corresponding to an age $>10^{10}$ years for a white dwarf of mass 1.0 -- 1.2 \Msun\ with a hydrogen atmosphere; \citealp{chabrier00a}), the apparent magnitude of a white dwarf companion in the $R$ band would be around 23.5, within reach of a relatively short observation. Future analysis will make use of additional optical and X--ray observations of \psr\ to definitively characterize the companion object and the evolutionary pathway which formed the system.


The astrometric information also allows us to calculate a number of corrections to the pulsar timing observables.  The  dominant contribution is the Shklovskii effect \citep{shklovskii70a}, where $\dot{P}_{\mathrm{Shk}}/P = \mu^{2}\mathrm{D}/\mathrm{c}$, where $\mu$ is the proper motion, D is the pulsar distance, c is the speed of light, and $\dot{P}$ is the pulsar spin period derivative.  Substituting the distance and velocity derived above, and taking $P$ as 32.82 ms \citep{boyles13a}, we obtain $\dot{P}_{\mathrm{Shk}} = \left(4.33 \pm 0.02\right) \times 10^{-20}$.  
The net effect of acceleration in the Galactic gravitational potential \citep[see e.g.,][]{nice95a} is negligible for \psr, less than 1\% of the Shklovskii effect.
The measured period derivative for \psr\ is $(4.74 \pm 0.03) \times 10^{-20}$, thus the intrinsic $\dot{P}$ is $\left(4.1 \pm 0.4\right) \times 10^{-21}$.  This revises the estimates of characteristic age $\tau_{\mathrm{c}}$ to 1.3 $\times$ 10$^{11}$ years and surface magnetic field strength $B_\mathrm{surf}$ to 3.7 $\times$ 10$^8$ G.  The very high value for $\tau_\mathrm{c}$ (the largest amongst known pulsars) confirms that the pulsar was only partially recycled.
 
 \section{The future of precision astrometry at 20 cm}
 \label{sec:general}

\subsection{The impact of calibrator structure evolution}

\begin{figure}
\begin{center}
\begin{tabular}{c}
\includegraphics[width=0.45\textwidth]{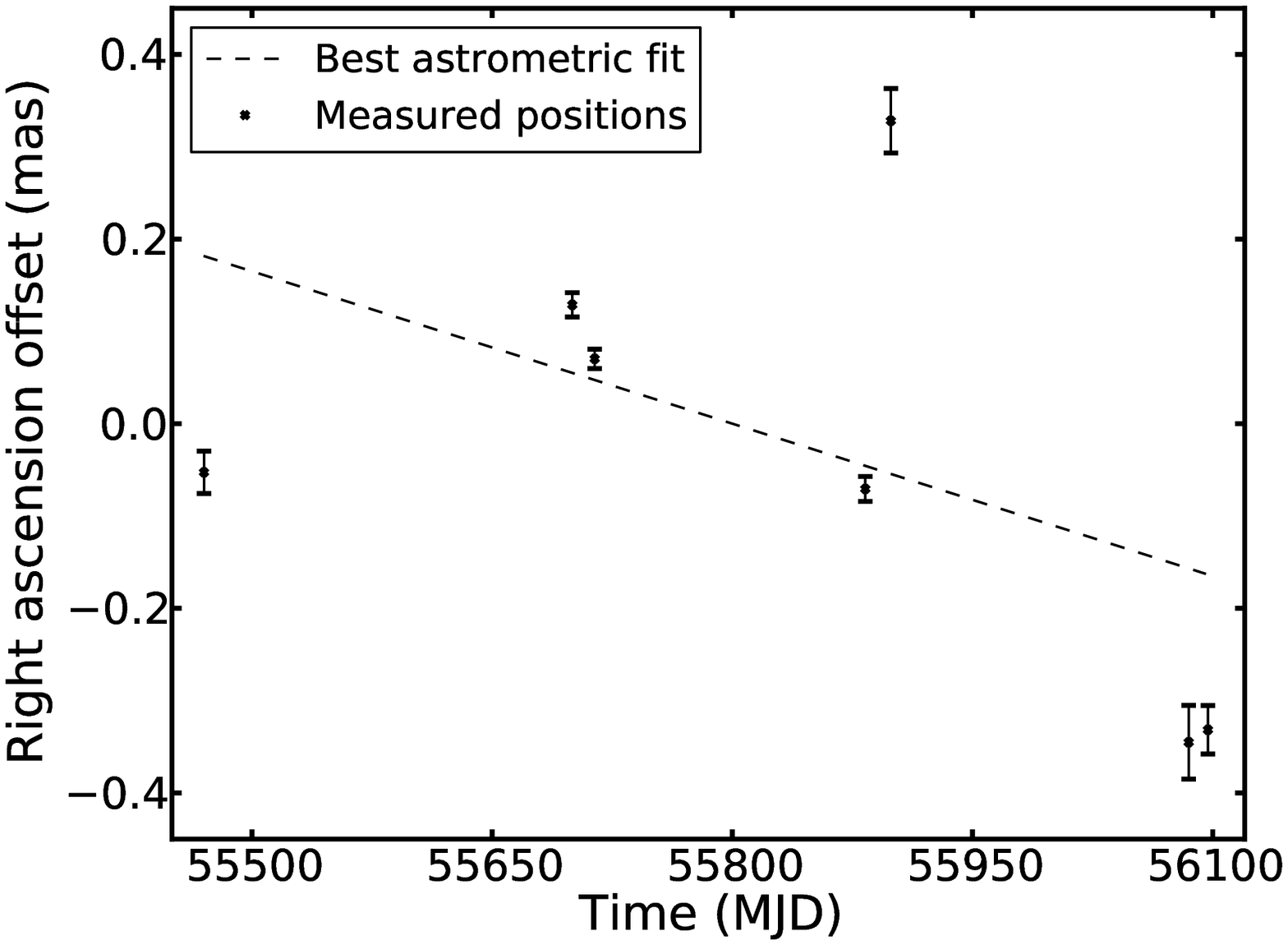} \\
\includegraphics[width=0.45\textwidth]{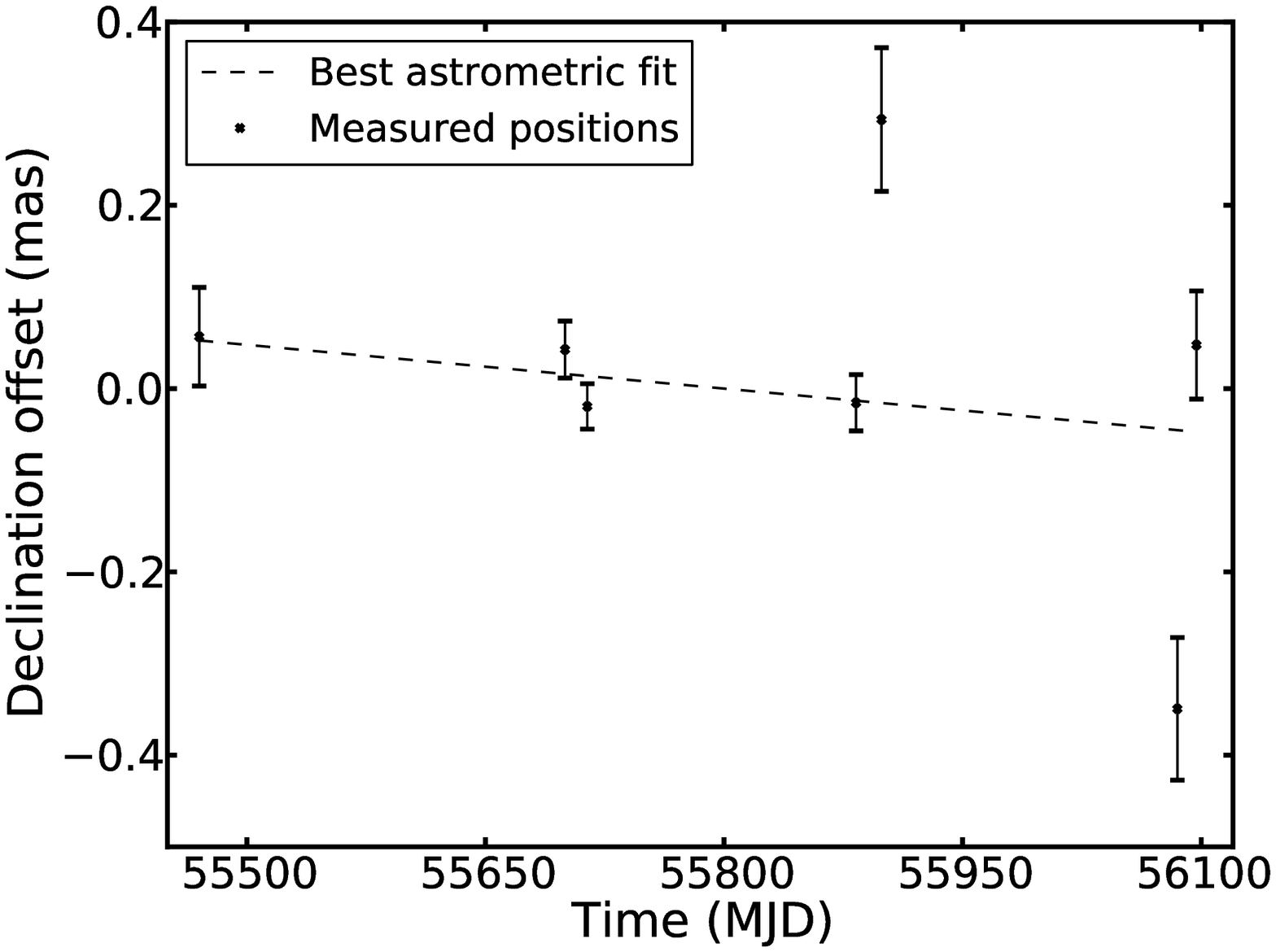} 
\end{tabular}
\end{center}
\caption{An astrometric fit to the position residuals for the secondary in--beam calibrator \ibtwo.  The top panel shows offset from the nominal position in right ascension, and the bottom panel shows offset in declination.}
\label{fig:ib2fit}
\end{figure}

The presence of a second in--beam calibrator, \ibtwo, affords us an opportunity to examine the potential contribution of calibrator structure evolution on astrometric accuracy.  In Figure~\ref{fig:ib2fit} we plot the astrometric position fits for \ibtwo, which should be consistent with a constant source position.   Over the 1.5 year observing period, however, the position centroid evolves markedly - particularly along the right ascension axis, where the deviations are more than an order of magnitude above the error bars.  As can be seen in Figure~\ref{fig:inbeams}, \ibtwo\ is clearly a core--jet system, and so significant structural evolution might be expected along the jet axis, which is almost exactly aligned with the right ascension axis.  The ejection of components along the jet axis which brighten, shift, and fade is almost certainly the major contributor to the position deviations.  A smaller portion of the apparent variation can also be ascribed to differing $uv$ coverage between the epochs, which will lead to position shifts if the model of the calibrator is imperfect (which is certainly the case, since the arcsecond scale flux is considerably greater than the total flux recovered in the VLBI image).  A final error component will be the differential atmosphere/ionosphere between \ibtwo\ and \ibone; the angular separation of these two sources is considerably larger than that between \ibone\ and \psr.  However, this contribution would not be expected to exceed $\sim$75 $\mu$as \citep{deller12b}.

Regardless of the exact ratio of the contributing sources of error, we conclude that the dominant impact comes from the fact that \ibtwo\ is a source with complicated and time--variable structure, and we can calculate the impact on astrometric accuracy in a hypothetical situation where it was the only calibrator available.  Transferring the position offsets from the fits to \ibone\ to the corresponding positions of \psr\ causes a dramatic reduction in quality -- the reduced $\chi^2$ of the fit to \psr\ would be 9, and the final parallax error balloons to over 30 $\mu$as.  This result highlights that while the focus to date for precision pulsar astrometry has been on obtaining sufficiently bright calibrators as close as possible to the target, careful attention should also be paid to morphological properties when selecting calibrators.  If at all possible, all viable calibrators should be obtained and results compared at the end of an astrometric campaign, enabling different sources of error to be estimated and ``traded off" for the best final result.
 
\subsection{Predicting astrometric precision}
\label{sec:predicting}

\begin{figure}
\begin{center}
\includegraphics*[height=0.45\textwidth]{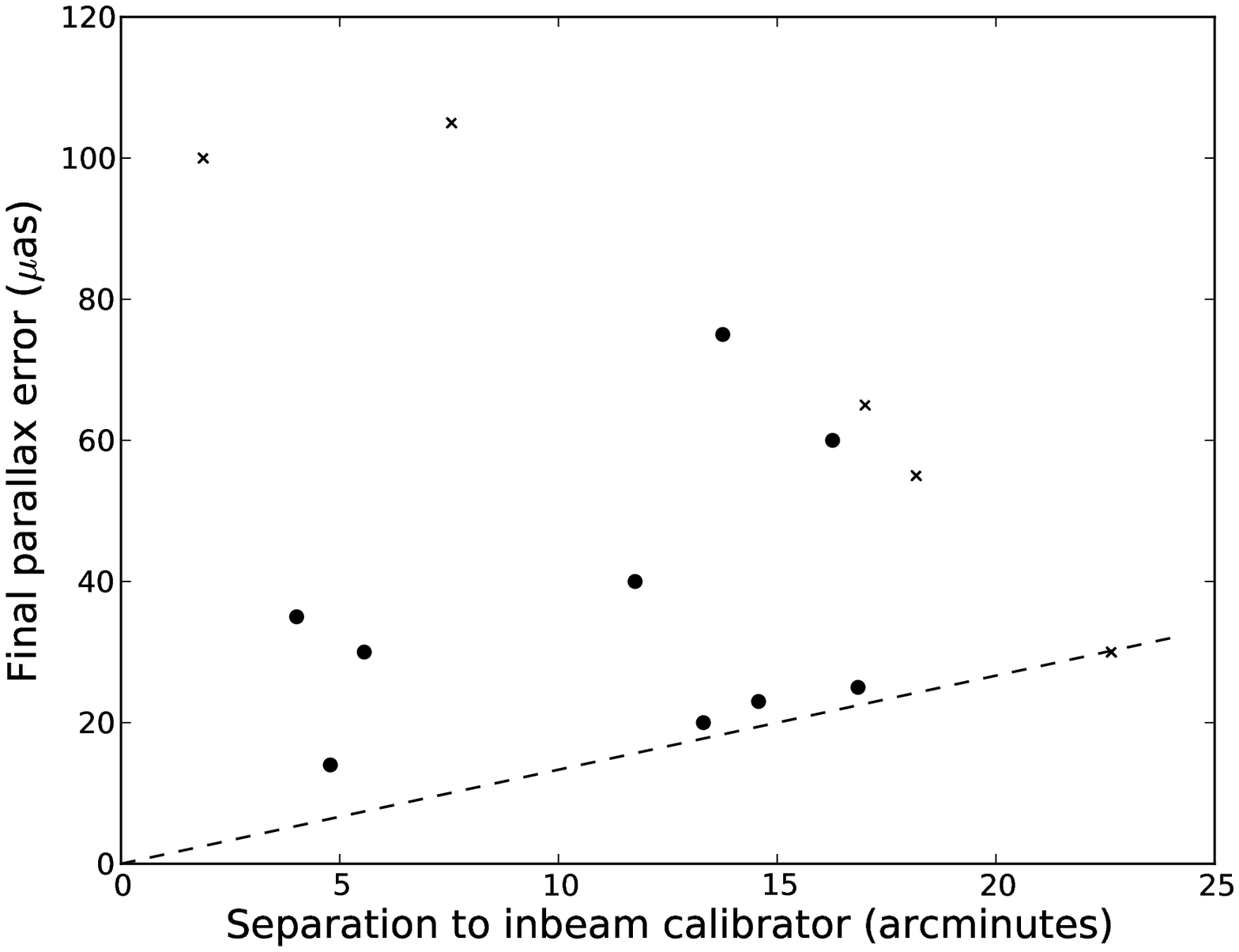}
\caption{Final parallax error plotted against separation to in--beam calibrators, for pulsars published in \citet{chatterjee09a}, \citet{deller12b} and this paper.  The astrometric fits for each pulsar used between 7 and 10 epochs.  Some of the calibrators used in \citet{chatterjee09a} were considerably fainter than desirable, such that the noise on the in--beam calibrator solutions is likely to be a dominant contributor to the total error budget.  These sources with questionable calibration (selected as those with peak flux density $<$ 9 mJy/beam; the 5 minute, 1$\sigma$ baseline sensitivity of the VLBA at the 256 Mbps recording rate used by \citealt{chatterjee09a} is 2.5 mJy) are marked on this plot with crosses; the remaining pulsars are marked with a filled circle.  The dashed line shows an empirical fit to the best attainable parallax accuracy with a given calibrator separation (equal to 1.33 $\mu$as per arcminute separation).  When pulsars which utilized low S/N calibrators are excluded, the trend towards larger errors at larger separations becomes visible, but it is also obvious that angular separation is rarely the sole contributing factor to astrometric accuracy.}
\label{fig:seperrors}
\end{center}
\end{figure}

Over the last several years, 15 pulsar parallaxes have been obtained using the VLBA at 1.6 GHz with in--beam calibrators \citep{chatterjee09a, deller12b}.  By the end of 2013, that number will increase five--fold, with the completion of the PSR$\pi$ program \citep{deller11b}.  It is therefore timely to take stock of the abilities and limitations of this method.  Figure~\ref{fig:seperrors} plots the final parallax error obtained for each pulsar against the angular separation to the (primary, if multiple were available) in--beam calibrator.  It is apparent that angular separation on its own is insufficient to predict attainable astrometric precision, as many sources with favorably small angular separations have relatively large errors.  In some of these cases (those plotted with an x symbol in Figure~\ref{fig:seperrors}) insufficient sensitivity on the calibrator is likely the reason.  For others, calibrator structure evolution such as that seen in \ibtwo\ may be at play.  In general, the random (radiometer noise) error in the target image does not contribute significantly to the error budget -- \psr\ is an exception in this regard.  This implies, of course, that more sensitive observations of \psr\ could lead to an even more accurate distance measurement.

Looking at only the best results as the angular separation increases shows a relatively constant linear trend with a parallax error of $\sim$1.33 $\mu$as per arcminute of calibrator--target separation, plotted as a dashed line on Figure~\ref{fig:seperrors}.  This represents a lower limit to the parallax error attainable in a typical VLBI observing campaign with $\sim$8 epochs under the observing conditions experienced to date.  Increasing the number of observing epochs could help reduce this further, but as the parallax error will only improve with the square root of the number of epochs (appropriately spaced in time), this can at best help by a factor of $\sim$2.  This guideline could prove useful in estimating accuracies for future astrometric campaigns.  However, as Figure~\ref{fig:seperrors} shows, separation alone is insufficient -- a calibrator must also be sufficiently bright and stable.  Accordingly, for any astrometric pulsar campaign it is useful to inspect all potential in--beam calibrators before commencing the campaign, and to make use of multiple calibrators if possible, even if the secondary and subsequent in--beam calibrators are at greater angular separations.

Finally, it is noteworthy that almost all of the observations shown in Figure~\ref{fig:seperrors} were made at a time closer to solar maximum than solar minimum -- the observations presented in \citet{chatterjee09a} took place between 2002 and 2005.  Since ionospheric activity is considerably higher at these times, it is reasonable to suppose that astrometric campaigns made closer to solar minimum could attain somewhat better results than the ``lower limit" presented above.

\subsection{Astrometry and pulsar timing arrays}

This project has shown that measurements of pulsar distances to sub--parsec accuracy are feasible.  In the future, it can be expected that the intersection of ultra--high--precision astrometric measurements with ultra--high--precision pulsar timing can lead to new probes of post--Newtonian physics.  Here, we consider the impact on one high--profile target -- long period gravitational waves, as measured by a pulsar timing array \citep[PTA;][]{hobbs10a}.  As shown by \citet{mingarella12a}, the addition of precision astrometric information allows the effect of gravitational wave emission by a binary supermassive black hole system on pulsar timing observables to be separated into components affecting the observer (on Earth) and the emitter (the pulsar, hundreds of parsecs distant).  A measurement of the pulsar term can immediately constrain the mass and spin of the two components of the black hole binary, providing information which is very difficult to infer by other (indirect) means.  The challenge is the required astrometric precision -- for a gravitational wave with a period of 12 years, a distance accurate to 0.4 pc is necessary to coherently connect the Earth and pulsar terms.  For \psr, this would require a factor--of--2 increase in the astrometric precision -- well within the realms of possibility given a concerted VLBA astrometric campaign.

However, the timing precision of \psr\ (residual rms 8 $\mu$s; \citealt{boyles13a}) is not currently high enough that it would add appreciable sensitivity to a PTA.  Around 40 pulsars are currently observed by PTA projects including the Parkes Pulsar Timing Array \citep[PPTA;][]{manchester13a}, NANOGrav \citep{demorest13a} and the European Pulsar Timing Array \citep[EPTA;][]{van-haasteren11a}, but just a handful of these are currently producing results at a precision sufficient to contribute significantly to the detection of gravitational waves.  The reference timing accuracy usually assumed for simulations of gravitational wave detection sensitivity is an rms residual of 100 ns \citep{jenet05a,verbiest09a}.  In the recent results published by \citet{manchester13a} and \citet{demorest13a}, less than 10 pulsars currently have residuals within a factor of 2 of this level (although recent EPTA results are not available, at most one or two additional sources at this level could be expected, given the high level of overlap between the PTA target lists).  

Of these high--accuracy pulsars, however, most are known or predicted to be far more distant than \psr.  As the parallax precision required for a given linear distance accuracy scales with the square of the distance, obtaining a distance accurate to 0.4 pc or better is beyond contemplation with current instrumentation for most potential targets.  Table~\ref{tab:ptapulsars} shows the distance to all high--precision PTA pulsars which are thought to be less than 1 kpc from the solar system, and the parallax accuracy required for a 0.4 pc distance error (of course, the predicted distances may be in error, making the task easier or harder than predicted). PSR J0437--4715 is the only potential target which is closer than \psr, and hence the only source requiring a less stringent level of astrometric precision.  However, the southern location of PSR J0437--4715 precludes observations with the VLBA, and while a high precision distance to PSR J0437--4715 has been obtained using the Long Baseline Array (LBA) in Australia, the heterogeneous nature of the LBA and the consequent small field of view of some of the elements makes the use of an in--beam calibrator virtually impossible, making it unlikely that LBA observations will be able to approach the accuracies seen with the VLBA.

\begin{deluxetable}{lcrcc}
\tabletypesize{\small}
\tablecaption{High--precision PTA pulsars (timing residuals $<$ 300 ns) predicted to be within 1 kpc of the solar system}
\tablewidth{0pt}
\tablehead{
\colhead{Pulsar} & \colhead{Predicted} & \colhead{Distance} & \colhead{Parallax } & \colhead{Required parallax accuracy} \\
\colhead{} & \colhead{distance (pc)} & \colhead{reference} & \colhead{signature (mas)} & \colhead{for $\Delta$d $<$ 0.4pc ($\mu$as)} 
}
\startdata
J0030+0451 & 240 & \citet{lommen06a} & 4.17 & 6.9 \\
J0437--4715 & 157 & \citet{deller08b} & 6.37	& 16.2 \\
J1744--1134 & 420 & \citet{verbiest09a} & 2.38	& 2.3 \\
J1857+0943 & 910 & \citet{cordes02a} &1.10 	& 0.5
\enddata
\label{tab:ptapulsars}
\end{deluxetable}


Table~\ref{tab:ptapulsars} suggests that for the foreseeable future (at least until the arrival of the second phase of the Square Kilometre Array), only PSR J0030+0451 and PSR J1744--1134 offer a credible hope of measuring a distance precisely enough to allow the investigation of individual gravitational wave sources using the pulsar term.  In each case, the best--case accuracy derived in Section~\ref{sec:predicting} predicts that a calibrator (or preferably more than one) within a few arcminutes of the target would be needed to reduce the systematic error contribution below the required threshold.  Within such a small radius, the brightest compact sources are likely to have a flux density $<$1 mJy/beam, which would demand the use of a very sensitive VLBI configuration (large telescopes and wide bandwidths).  However, even if it proves impossible to obtain the desired accuracy on these two candidates, other possibilities still exist -- current and future surveys may yet discover additional nearby MSPs suitable for PTA observations, or improvements in pulsar timing precision bring current, nearby PTA pulsars into the timing accuracy range where VLBI distance measurements become appealing.

\acknowledgements  The National Radio Astronomy Observatory is a facility of the National Science Foundation operated under cooperative agreement by Associated Universities, Inc.  ATD was supported by an NWO Veni Fellowship.  Pulsar research at UBC is supported by an NSERC Discovery Grant. The authors thank S Chatterjee for providing data products from earlier VLBA astrometric observations and J. Moldon for useful discussions regarding orbital fitting.  We thank the referee for providing valuable feedback which improved this manuscript.

\bibliographystyle{apj}
\bibliography{deller_thesis}

\begin{thebibliography}{}
\expandafter\ifx\csname natexlab\endcsname\relax\def\natexlab#1{#1}\fi

\bibitem[{{Becker} {et~al.}(1995){Becker}, {White}, \& {Helfand}}]{becker95a}
{Becker}, R.~H., {White}, R.~L., \& {Helfand}, D.~J. 1995, \apj, 450, 559

\bibitem[{{Boyles} {et~al.}(2011){Boyles}, {Lorimer}, {McLaughlin}, {Ransom},
  {Lynch}, {Kaspi}, {Archibald}, {Stairs}, {McPhee}, {Roberts}, {Kondratiev},
  {Hessels}, {van Leeuwen}, {Champion}, {Deller}, \& {Dunlap}}]{boyles11a}
{Boyles}, J., {Lorimer}, D.~R., {McLaughlin}, M.~A., {et~al.} 2011, in American
  Institute of Physics Conference Series, Vol. 1357, Radio Pulsars: An
  Astrophysical Key to Unlock the Secrets of the Universe, ed. M.~{Burgay},
  N.~{D'Amico}, P.~{Esposito}, A.~{Pellizzoni}, \& A.~{Possenti}, 32--35

\bibitem[{{Boyles} {et~al.}(2013){Boyles}, {Lynch}, {Ransom}, {Stairs},
  {Lorimer}, {McLaughlin}, {Hessels}, {Kaspi}, {Kondratiev}, {Archibald},
  {Berndsen}, {Cardoso}, {Cherry}, {Epstein}, {Karako-Argaman}, {McPhee},
  {Pennucci}, {Roberts}, {Stovall}, \& {van Leeuwen}}]{boyles13a}
{Boyles}, J., {Lynch}, R.~S., {Ransom}, S.~M., {et~al.} 2013, \apj, 763, 80

\bibitem[{{Camilo} {et~al.}(2001){Camilo}, {Lyne}, {Manchester}, {Bell},
  {Stairs}, {D'Amico}, {Kaspi}, {Possenti}, {Crawford}, \& {McKay}}]{camilo01a}
{Camilo}, F., {Lyne}, A.~G., {Manchester}, R.~N., {et~al.} 2001, \apjl, 548,
  L187

\bibitem[{{Chabrier} {et~al.}(2000){Chabrier}, {Brassard}, {Fontaine}, \&
  {Saumon}}]{chabrier00a}
{Chabrier}, G., {Brassard}, P., {Fontaine}, G., \& {Saumon}, D. 2000, \apj,
  543, 216

\bibitem[{{Chatterjee} {et~al.}(2009){Chatterjee}, {Brisken}, {Vlemmings},
  {Goss}, {Lazio}, {Cordes}, {Thorsett}, {Fomalont}, {Lyne}, \&
  {Kramer}}]{chatterjee09a}
{Chatterjee}, S., {Brisken}, W.~F., {Vlemmings}, W.~H.~T., {et~al.} 2009, \apj,
  698, 250

\bibitem[{{Chaurasia} \& {Bailes}(2005)}]{chaurasia05a}
{Chaurasia}, H.~K., \& {Bailes}, M. 2005, \apj, 632, 1054

\bibitem[{{Cordes} \& {Lazio}(2002)}]{cordes02a}
{Cordes}, J.~M., \& {Lazio}, T.~J.~W. 2002, ArXiv e-prints, 0207156,
  astro-ph/0207156

\bibitem[{{Deller} {et~al.}(2009){Deller}, {Tingay}, {Bailes}, \&
  {Reynolds}}]{deller09b}
{Deller}, A.~T., {Tingay}, S.~J., {Bailes}, M., \& {Reynolds}, J.~E. 2009,
  \apj, 701, 1243

\bibitem[{{Deller} {et~al.}(2007){Deller}, {Tingay}, {Bailes}, \&
  {West}}]{deller07a}
{Deller}, A.~T., {Tingay}, S.~J., {Bailes}, M., \& {West}, C. 2007, \pasp, 119,
  318

\bibitem[{{Deller} {et~al.}(2008){Deller}, {Verbiest}, {Tingay}, \&
  {Bailes}}]{deller08b}
{Deller}, A.~T., {Verbiest}, J.~P.~W., {Tingay}, S.~J., \& {Bailes}, M. 2008,
  \apjl, 685, L67

\bibitem[{{Deller} {et~al.}(2011{\natexlab{a}}){Deller}, {Brisken}, {Phillips},
  {Morgan}, {Alef}, {Cappallo}, {Middelberg}, {Romney}, {Rottmann}, {Tingay},
  \& {Wayth}}]{deller11a}
{Deller}, A.~T., {Brisken}, W.~F., {Phillips}, C.~J., {et~al.}
  2011{\natexlab{a}}, \pasp, 123, 275

\bibitem[{{Deller} {et~al.}(2011{\natexlab{b}}){Deller}, {Brisken},
  {Chatterjee}, {Cordes}, {Goss}, {Janssen}, {Kovalev}, {Lazio}, {Petrov}, \&
  {Stappers}}]{deller11b}
{Deller}, A.~T., {Brisken}, W.~F., {Chatterjee}, S., {et~al.}
  2011{\natexlab{b}}, in Proceedings of the 20th EVGA Meeting, held 29-31
  March, 2011 at Max-Planck-Institut f{\"u}r Radioastronomie, Bonn, Germany.
  Edited by Walter Alef, Simone Bernhart, and Axel Nothnagel, p.178, ed.
  W.~{Alef}, S.~{Bernhart}, \& A.~{Nothnagel}, 178

\bibitem[{{Deller} {et~al.}(2012){Deller}, {Archibald}, {Brisken},
  {Chatterjee}, {Janssen}, {Kaspi}, {Lorimer}, {Lyne}, {McLaughlin}, {Ransom},
  {Stairs}, \& {Stappers}}]{deller12b}
{Deller}, A.~T., {Archibald}, A.~M., {Brisken}, W.~F., {et~al.} 2012, \apjl,
  756, L25

\bibitem[{{Demorest} {et~al.}(2013){Demorest}, {Ferdman}, {Gonzalez}, {Nice},
  {Ransom}, {Stairs}, {Arzoumanian}, {Brazier}, {Burke-Spolaor}, {Chamberlin},
  {Cordes}, {Ellis}, {Finn}, {Freire}, {Giampanis}, {Jenet}, {Kaspi}, {Lazio},
  {Lommen}, {McLaughlin}, {Palliyaguru}, {Perrodin}, {Shannon}, {Siemens},
  {Stinebring}, {Swiggum}, \& {Zhu}}]{demorest13a}
{Demorest}, P.~B., {Ferdman}, R.~D., {Gonzalez}, M.~E., {et~al.} 2013, \apj,
  762, 94

\bibitem[{{Efron} \& {Tibshirani}(1991)}]{efron91a}
{Efron}, B., \& {Tibshirani}, R. 1991, Science, 253, 390

\bibitem[{{Hobbs} {et~al.}(2010){Hobbs}, {Archibald}, {Arzoumanian}, {Backer},
  {Bailes}, {Bhat}, {Burgay}, {Burke-Spolaor}, {Champion}, {Cognard}, {Coles},
  {Cordes}, {Demorest}, {Desvignes}, {Ferdman}, {Finn}, {Freire}, {Gonzalez},
  {Hessels}, {Hotan}, {Janssen}, {Jenet}, {Jessner}, {Jordan}, {Kaspi},
  {Kramer}, {Kondratiev}, {Lazio}, {Lazaridis}, {Lee}, {Levin}, {Lommen},
  {Lorimer}, {Lynch}, {Lyne}, {Manchester}, {McLaughlin}, {Nice}, {Oslowski},
  {Pilia}, {Possenti}, {Purver}, {Ransom}, {Reynolds}, {Sanidas}, {Sarkissian},
  {Sesana}, {Shannon}, {Siemens}, {Stairs}, {Stappers}, {Stinebring},
  {Theureau}, {van Haasteren}, {van Straten}, {Verbiest}, {Yardley}, \&
  {You}}]{hobbs10a}
{Hobbs}, G., {Archibald}, A., {Arzoumanian}, Z., {et~al.} 2010, Classical and
  Quantum Gravity, 27, 084013

\bibitem[{{Jenet} {et~al.}(2005){Jenet}, {Hobbs}, {Lee}, \&
  {Manchester}}]{jenet05a}
{Jenet}, F.~A., {Hobbs}, G.~B., {Lee}, K.~J., \& {Manchester}, R.~N. 2005,
  \apjl, 625, L123

\bibitem[{{Kettenis} {et~al.}(2006){Kettenis}, {van Langevelde}, {Reynolds}, \&
  {Cotton}}]{kettenis06a}
{Kettenis}, M., {van Langevelde}, H.~J., {Reynolds}, C., \& {Cotton}, B. 2006,
  in Astronomical Society of the Pacific Conference Series, Vol. 351,
  Astronomical Data Analysis Software and Systems XV, ed. C.~{Gabriel},
  C.~{Arviset}, D.~{Ponz}, \& S.~{Enrique}, 497

\bibitem[{{Loinard} {et~al.}(2007){Loinard}, {Torres}, {Mioduszewski},
  {Rodr{\'{\i}}guez}, {Gonz{\'a}lez-L{\'o}pezlira}, {Lachaume}, {V{\'a}zquez},
  \& {Gonz{\'a}lez}}]{loinard07a}
{Loinard}, L., {Torres}, R.~M., {Mioduszewski}, A.~J., {et~al.} 2007, \apj,
  671, 546

\bibitem[{{Lommen} {et~al.}(2006){Lommen}, {Kipphorn}, {Nice}, {Splaver},
  {Stairs}, \& {Backer}}]{lommen06a}
{Lommen}, A.~N., {Kipphorn}, R.~A., {Nice}, D.~J., {et~al.} 2006, \apj, 642,
  1012

\bibitem[{{Lynch} {et~al.}(2013){Lynch}, {Boyles}, {Ransom}, {Stairs},
  {Lorimer}, {McLaughlin}, {Hessels}, {Kaspi}, {Kondratiev}, {Archibald},
  {Berndsen}, {Cardoso}, {Cherry}, {Epstein}, {Karako-Argaman}, {McPhee},
  {Pennucci}, {Roberts}, {Stovall}, \& {van Leeuwen}}]{lynch13a}
{Lynch}, R.~S., {Boyles}, J., {Ransom}, S.~M., {et~al.} 2013, \apj, 763, 81

\bibitem[{{Lyne} {et~al.}(2004){Lyne}, {Burgay}, {Kramer}, {Possenti},
  {Manchester}, {Camilo}, {McLaughlin}, {Lorimer}, {D'Amico}, {Joshi},
  {Reynolds}, \& {Freire}}]{lyne04a}
{Lyne}, A.~G., {Burgay}, M., {Kramer}, M., {et~al.} 2004, Science, 303, 1153

\bibitem[{{Manchester} {et~al.}(2013){Manchester}, {Hobbs}, {Bailes}, {Coles},
  {van Straten}, {Keith}, {Shannon}, {Bhat}, {Brown}, {Burke-Spolaor},
  {Champion}, {Chaudhary}, {Edwards}, {Hampson}, {Hotan}, {Jameson}, {Jenet},
  {Kesteven}, {Khoo}, {Kocz}, {Maciesiak}, {Oslowski}, {Ravi}, {Reynolds},
  {Sarkissian}, {Verbiest}, {Wen}, {Wilson}, {Yardley}, {Yan}, \&
  {You}}]{manchester13a}
{Manchester}, R.~N., {Hobbs}, G., {Bailes}, M., {et~al.} 2013, \pasa, 30, 17

\bibitem[{Mingarelli {et~al.}(2012)Mingarelli, Grover, Sidery, Smith, \&
  Vecchio}]{mingarella12a}
Mingarelli, C. M.~F., Grover, K., Sidery, T., Smith, R. J.~E., \& Vecchio, A.
  2012, Phys. Rev. Lett., 109, 081104

\bibitem[{{Mold{\'o}n} {et~al.}(2011){Mold{\'o}n}, {Johnston}, {Rib{\'o}},
  {Paredes}, \& {Deller}}]{moldon11a}
{Mold{\'o}n}, J., {Johnston}, S., {Rib{\'o}}, M., {Paredes}, J.~M., \&
  {Deller}, A.~T. 2011, \apjl, 732, L10

\bibitem[{{Mold{\'o}n} {et~al.}(2012){Mold{\'o}n}, {Rib{\'o}}, {Paredes},
  {Brisken}, {Dhawan}, {Kramer}, {Lyne}, \& {Stappers}}]{moldon12a}
{Mold{\'o}n}, J., {Rib{\'o}}, M., {Paredes}, J.~M., {et~al.} 2012, \aap, 543,
  A26

\bibitem[{{Nagayama} {et~al.}(2011){Nagayama}, {Omodaka}, {Handa}, {Honma},
  {Kobayashi}, {Kawaguchi}, \& {Ueno}}]{nagayama11a}
{Nagayama}, T., {Omodaka}, T., {Handa}, T., {et~al.} 2011, \pasj, 63, 719

\bibitem[{{Nice} \& {Taylor}(1995)}]{nice95a}
{Nice}, D.~J., \& {Taylor}, J.~H. 1995, \apj, 441, 429

\bibitem[{{Reid} {et~al.}(2009){Reid}, {Menten}, {Zheng}, {Brunthaler},
  {Moscadelli}, {Xu}, {Zhang}, {Sato}, {Honma}, {Hirota}, {Hachisuka}, {Choi},
  {Moellenbrock}, \& {Bartkiewicz}}]{reid09a}
{Reid}, M.~J., {Menten}, K.~M., {Zheng}, X.~W., {et~al.} 2009, \apj, 700, 137

\bibitem[{{Shepherd}(1997)}]{shepherd97a}
{Shepherd}, M.~C. 1997, in Astronomical Society of the Pacific Conference
  Series, Vol. 125, Astronomical Data Analysis Software and Systems VI, ed.
  G.~{Hunt} \& H.~{Payne}, 77

\bibitem[{{Shklovskii}(1970)}]{shklovskii70a}
{Shklovskii}, I.~S. 1970, Soviet Astronomy, 13, 562

\bibitem[{{Splaver} {et~al.}(2005){Splaver}, {Nice}, {Stairs}, {Lommen}, \&
  {Backer}}]{splaver05a}
{Splaver}, E.~M., {Nice}, D.~J., {Stairs}, I.~H., {Lommen}, A.~N., \& {Backer},
  D.~C. 2005, \apj, 620, 405

\bibitem[{{van Haasteren} {et~al.}(2011){van Haasteren}, {Levin}, {Janssen},
  {Lazaridis}, {Kramer}, {Stappers}, {Desvignes}, {Purver}, {Lyne}, {Ferdman},
  {Jessner}, {Cognard}, {Theureau}, {D'Amico}, {Possenti}, {Burgay},
  {Corongiu}, {Hessels}, {Smits}, \& {Verbiest}}]{van-haasteren11a}
{van Haasteren}, R., {Levin}, Y., {Janssen}, G.~H., {et~al.} 2011, \mnras, 414,
  3117

\bibitem[{{Verbiest} {et~al.}(2008){Verbiest}, {Bailes}, {van Straten},
  {Hobbs}, {Edwards}, {Manchester}, {Bhat}, {Sarkissian}, {Jacoby}, \&
  {Kulkarni}}]{verbiest08a}
{Verbiest}, J.~P.~W., {Bailes}, M., {van Straten}, W., {et~al.} 2008, \apj,
  679, 675

\bibitem[{{Verbiest} {et~al.}(2009){Verbiest}, {Bailes}, {Coles}, {Hobbs}, {van
  Straten}, {Champion}, {Jenet}, {Manchester}, {Bhat}, {Sarkissian}, {Yardley},
  {Burke-Spolaor}, {Hotan}, \& {You}}]{verbiest09a}
{Verbiest}, J.~P.~W., {Bailes}, M., {Coles}, W.~A., {et~al.} 2009, \mnras, 400,
  951

\end{thebibliography}
\end{document}